\documentclass[graybox]{svmult}

\usepackage{mathptmx}       
\usepackage{helvet}         
\usepackage{courier}        
\usepackage{type1cm}        
\usepackage{makeidx}         
\usepackage{graphicx}        
\usepackage{multicol}        
\usepackage[bottom]{footmisc}

\makeindex             

\begin{document}

\title*{Compact Stars - How Exotic Can They Be?}
\author{S. Schramm, V. Dexheimer, R. Negreiros, J. Steinheimer, T. Sch\"urhoff}
\institute{S. Schramm \at FIAS, JW Goethe -Universit\"at, Ruth Moufang - Str. 1, D-60438 Frankfurt am Main, Germany \email{schramm@fias.uni-frankfurt.de}}
%
%
\maketitle

\abstract{
Strong interaction physics under extreme conditions of high temperature and/or density is of central interest in modern nuclear physics for experimentalists and theorists alike.
In order to investigate such systems, model approaches that include hadrons and quarks in a unified approach, will be discussed. Special attention will be given to high-density matter as it occurs in neutron stars.
Given the current observational limits for neutron star masses, the properties of hyperonic and hybrid stars will be determined. In this context especially the question of the extent, to which exotic particles like hyperons and quarks affect star masses, will be discussed.
}

\section{Introduction}

\label{sec:1}
Currently, the only way to investigate the properties of extremely dense and relatively cold strongly interacting matter is the study of the properties of neutron stars, or more general, compact stars. Depending on the equation of state, the central density of such an object might reach or even exceed the value of ten times the ground state density of nuclear matter.
In this sense, neutron star physics complements the efforts by many ultra-relativistic heavy-ion experiments to create a hot fireball in the collision, which, depending on the collision energy, contain varying amounts of net baryonic density.
Both of these areas can be comprised as investigation of strong interaction physics for hot and/or dense systems.
In order to connect and relate results from the different regimes of density and temperature, models that can realistically describe all those conditions have to be developed and studied. In particular, this includes a quantitatively reasonable description of ground state nuclear matter and finite nuclei, as well as the correct asymptotic states at large values of temperature and density, i. e. quarks and gluons.
Such an approach has been developed and refined over the recent years. In this article we will study some general observations within this theoretical approach related to the possibility of exotic matter in a neutron star, including hyperons, kaon condensates and quarks.

A main problem of extracting the equation of state of very dense matter via the study of neutron stars is that this matter is in the core of the star and, thus, not directly detectable (with the exception of neutrinos that can leave the star without secondary interactions). For theoretical modeling one therefore has to compare model results with the observational data available, including star masses, radii, temperature/cooling behavior, and rotational periods. Here, there are still relatively few data that can be used for cooling studies and the radii are very difficult to pin down to an accuracy of about 1 km or less. As a consequence, the main input for modeling is still the mass of the star. This is especially true, due to the recent accurate measurements of two heavy neutron stars 
 PSR J1614-2230 of $M = 1.97 \pm 0.04\, M_\odot$ \cite{197} and PSR J0348-0432 with a value of $M = 2.01 \pm 0.04\,M_\odot$ \cite{Antoniadis:2013pzd}.
These high values point to a  stiff equation of state at high densities. How to reconcile these numbers with theory will be discussed in the following. 

\section{The Hadronic Model}
\label{sec:2}

Within a simple (SU(2) isospin) description of a neutron star, the particles occurring in the cold star are neutrons, protons, and electrons. 
Studying the early, proto-neutron star phase, one also has to take into account neutrinos. As the new benchmark stellar masses are rather high,
the best scenario is to have relatively few populated degrees of freedom  inside of the star, as in this case filling up the energy levels of the particles leads to high fermi momenta, which in turn translates to high pressure and a stiff equation of state (EoS). Such a stiff EoS can support larger star masses against gravitational collapse than a soft EoS, therefore leading to larger maximum star masses. However, from the observation of many hyper nuclei it is well established that at least the Lambda hyperon is bound in nuclear matter with a binding energy of about 30 MeV. There are indications of binding of Xi baryons in nuclear matter with a binding of around 20 MeV \cite{Dover:1982ng}. The situation of the Sigma hyperons is less clear, however, scattering and Sigma-atom data point to a repulsive potential of the Sigma \cite{Harada:2005hs}. Overall a realistic model for the description of dense (and hot) matter has to contain hyperons, i. e., should be constructed within a flavor-SU(3) scheme.

We have followed this path for a number of years, developing an effective chiral SU(3) hadronic model, in which the baryonic masses are largely generated by their coupling to scalar fields that attain a vacuum expectation value due to spontaneous symmetry breaking, analogously to simple chiral $\sigma$ models.

The model is described in detail in various publications \cite{Papazoglou:1997uw,Papazoglou:1998vr,deformed}.
The key relations of the model are as follows.
The effective baryon masses $m_i^*$ read
\begin{equation}
m_i^* = g_{i\sigma} \sigma + g_{i\zeta} \zeta + g_{i\delta} \delta + \delta m_i ~~.
\label{mass}
\end{equation}
The masses are generated by couplings to the scalar fields ($\sigma, \delta, \zeta$) whose expectation values are related to the scalar quark condensates, i. e.
$\sigma \sim <\overline{u}u + \overline{d}d> $,  
$\zeta \sim <\overline{s}s>$, and
$\delta \sim <\overline{u}u - \overline{d}d> $.
A mass term $\delta m_i$ breaks chiral and SU(3) symmetry explicitly. The various couplings contained in the equation result from the SU(3) coupling scheme.
Thus, the non-zero vacuum expectation values of $\sigma$ and $\zeta$ generate the baryonic masses in the vacuum, while the change of the scalar fields at finite density or temperature are responsible for the scalar attraction. 
The other crucial relation defines the effective baryonic chemical potentials
\begin{equation}
\tilde{\mu}_i = \mu_i - g_{i\omega} \omega - g_{i\rho} \rho - g_{i\phi} \phi  ~~,
\label{cp}
\end{equation}
where the different fields $\omega, \rho$ and $\phi$ are the analogous vector fields to the $\sigma, \delta, $ and $\zeta$, respectively.  The fields acquire non-zero mean fields in dense matter and, therefore, shift the effective chemical potentials of the particles.
In the case of the isospin 0 fields $\omega$ and $\phi$ this yields to a reduction of the chemical potential, which corresponds to a repulsive interaction, whereas for the $\rho$ the effect depends on the
isospin of the baryon.
\begin{figure}[th]
\centerline{\includegraphics[width=6cm]{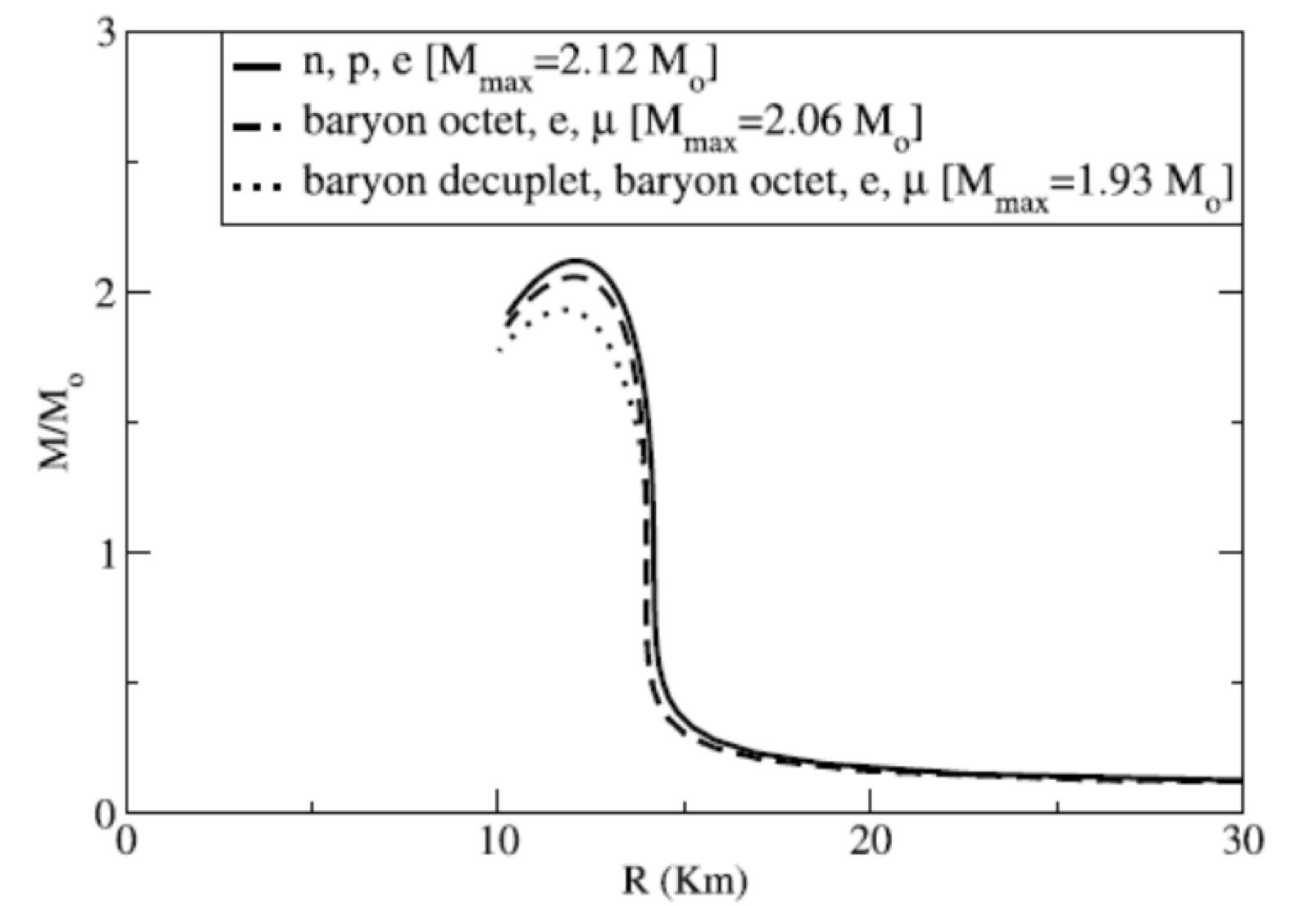}\includegraphics[width=6cm]{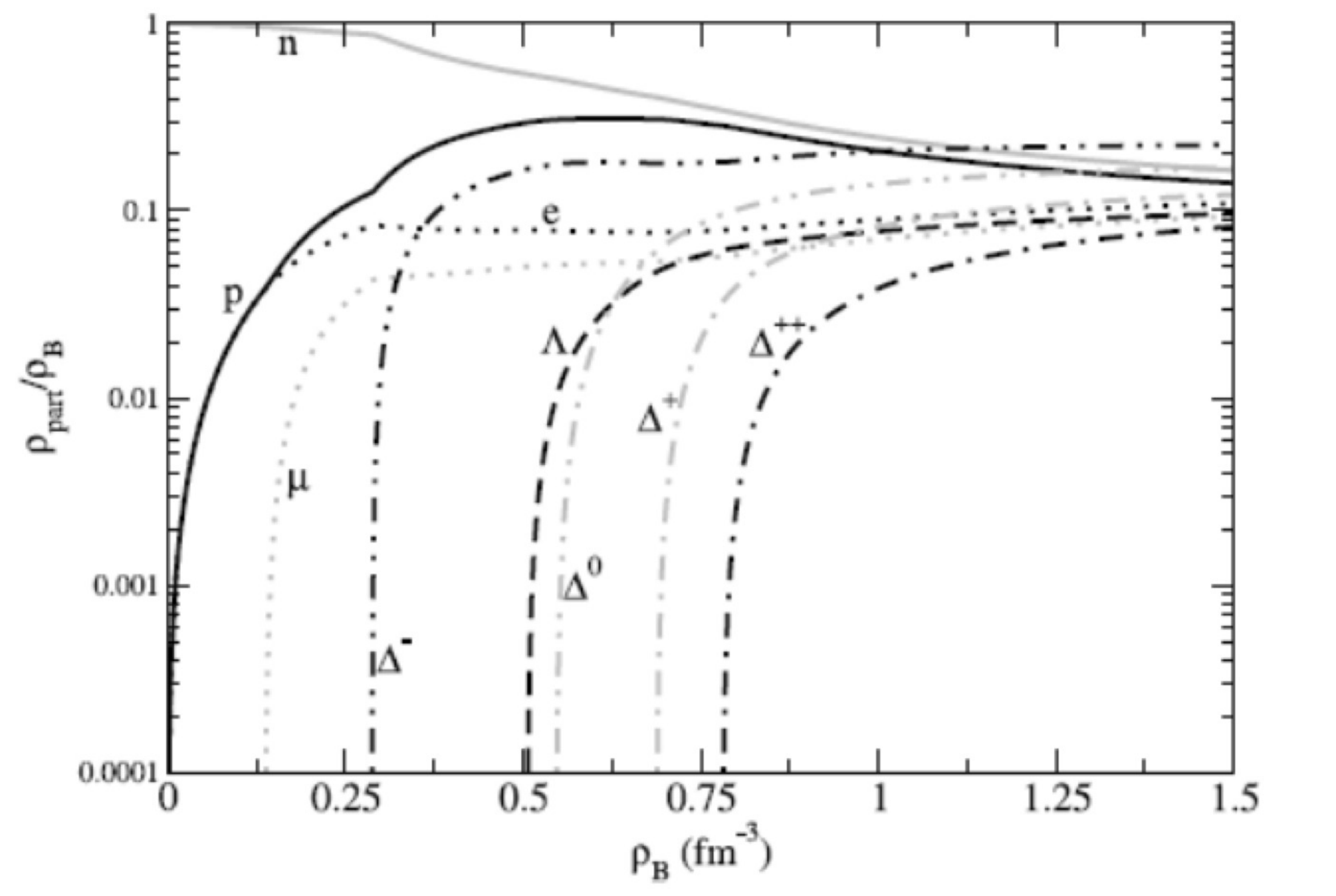}}
\vspace*{5pt}
\caption{Left: Mass-radius diagram for compact stars in the chiral SU(3) model. The curves distinguish results for nucleonic particles only, nucleons and hyperons, and nucleons, hyperons, and $\Delta$ resonances, respectively.
Right: Normalized particle densities in the star including the $\Delta$ resonances. $\Delta$'s largely replace hyperons inside the star \protect\cite{Dexheimer:2008ax}. } 
\label{star}
\end{figure}

\begin{figure}[th]
\centerline{\includegraphics[width=12cm,height=5cm]{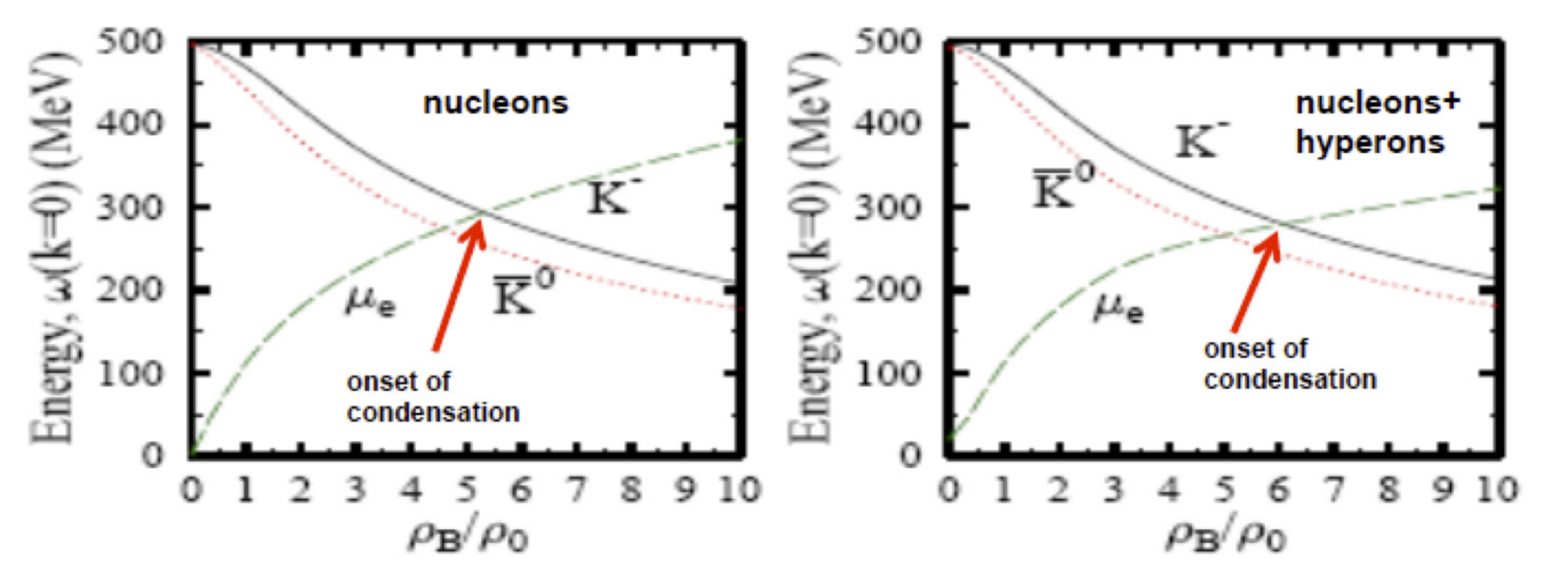}}
\vspace*{5pt}
\caption{Energy of $K^-$ and $\overline{K}^0$  at rest as function of density. When the energy of the $K^-$ drops below the electron chemical potential kaon condensation sets in. The left panel shows the result for nucleonic star matter, the right panel includes also hyperons \protect\cite{Mishra:2009bp}.}
\label{kaons}
\end{figure}

\section{Results}
\label{sec:3}
With these preliminaries, we can address the topic of this article, the influence of exotic particles on the compact star solutions. Here, for the
sake of simplicity, we will only consider static spherical stars, which can be solved straightforwardly by integrating the corresponding Tolman-Oppenheimer-Volkoff equations \cite{tov1,tov2} . Performing such a calculation for the purely hadronic case (and using the star-optimized coupling parameters) we obtain masses and radii of the stars as shown in the left panel of Fig. \ref{star} \cite{Dexheimer:2008ax}. The figure distinguishes the cases for a purely nucleonic star, a star with hyperons, i.e. a hyper star, as well as solutions when in addition the baryon decuplet is included, which is synonymous to adding $\Delta$ baryons as the other states in the decuplet are too heavy to be populated.
One can see that adding new degrees of freedom leads to a reduction of the maximum star mass, however, all scenarios are in good agreement with current observations.
The right panel of Fig. 1 shows that the structure of a nucleonic star can be changed quite drastically by including $\Delta$ baryons.
The fact that the hyper star masses do not drop considerably compared to the nucleonic case originates from the fact that the model also includes the repulsive $\phi$ meson exchange between hyperons, which suppresses their densities.

In addition to adding strangeness to the system by including hyperons, a condensation of $K^-$ mesons in very dense matter is in principle possible \cite{Kaplan:1986yq}.
This results from the fact that a $K^-$ consists of an anti-up quark and a strange quark. For the anti-quark the vector repulsion experienced by the baryons changes to a strong attraction. On the other hand the s quark in the meson does not couple strongly to the matter as long as there are not too many hyperons in the system, in sum leading to a strong attraction for the $K^-$  in matter.
There is a substantial amount of electrons (and some muons) in the star as the whole system has to be electrically neutral. In case the electron chemical potential becomes larger than the energy of a $K^-$ at rest, the system will start to be neutralized by $K^-$ instead of additional $e^-$. As the kaons are bosons, they can condense in the same zero-momentum state. This in turn leads to a strong reduction of the pressure of the system (no kinetic energy for the kaons), softening the equation of state considerably and therefore reducing star masses. However, a calculation of the effective kaon masses within the model shows that the condensation effect
only shows up at densities of about $6\,\rho_0$ and beyond, which are not reached in the center of the star within our model (Fig. \ref{kaons}) \cite{Mishra:2009bp,Mishra:2008kg}. Thus, kaon condensation does not affect the previously obtained results.

\section{The Effect of Quarks}
\label{sec:4}

\begin{figure}[th]
\centerline{\includegraphics[width=6cm]{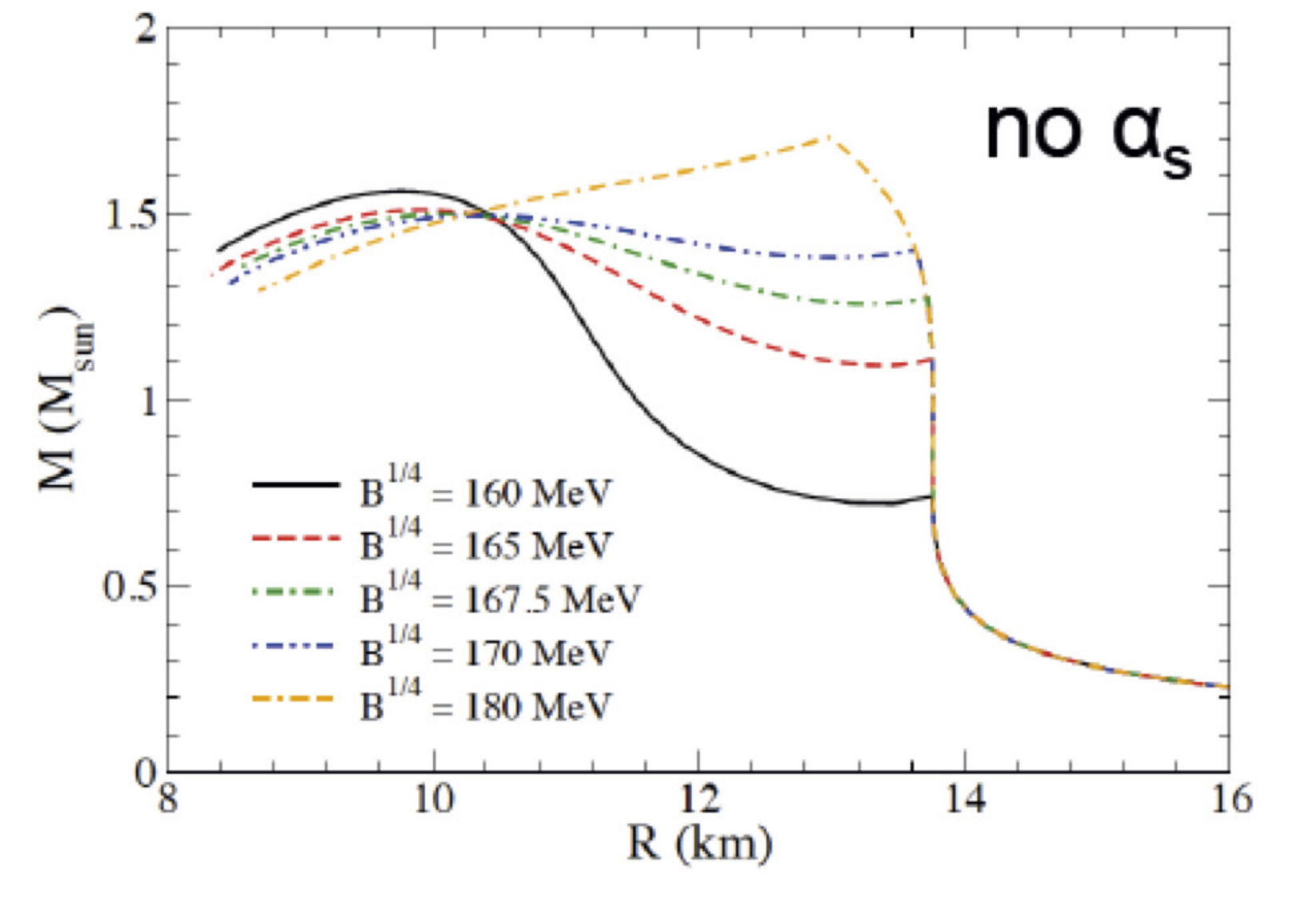}\includegraphics[width=6cm]{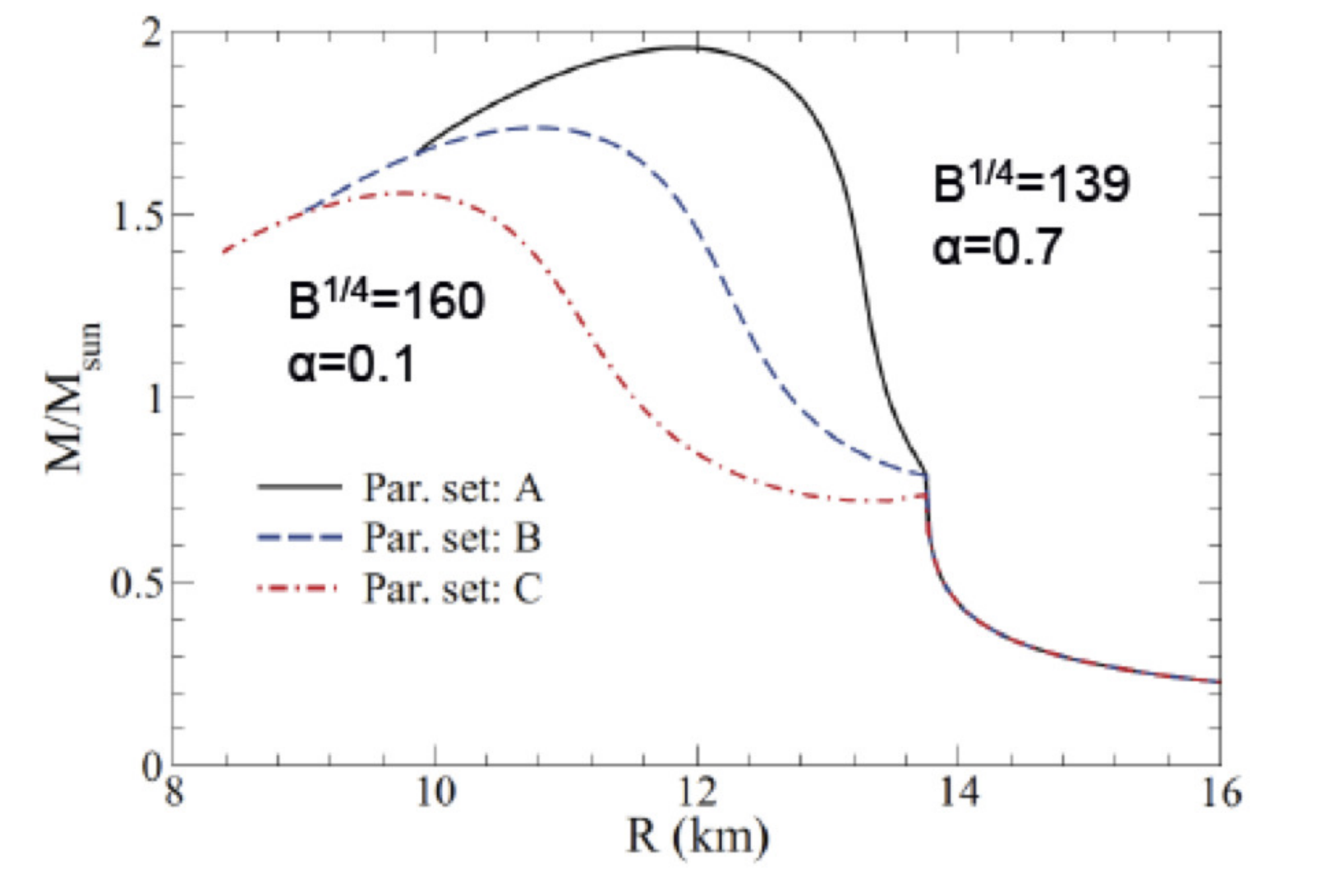}}
\vspace*{5pt}
\caption{Left: Hybrid star mass-radius diagram for different choices of the vacuum pressure $B$ are shown (compare Fig. \ref{MIT}).
Right: Similar calculation as on the left, including quark interactions.}
\label{MIT}
\end{figure}

\begin{figure}[th]
\centerline{\includegraphics[width=9cm]{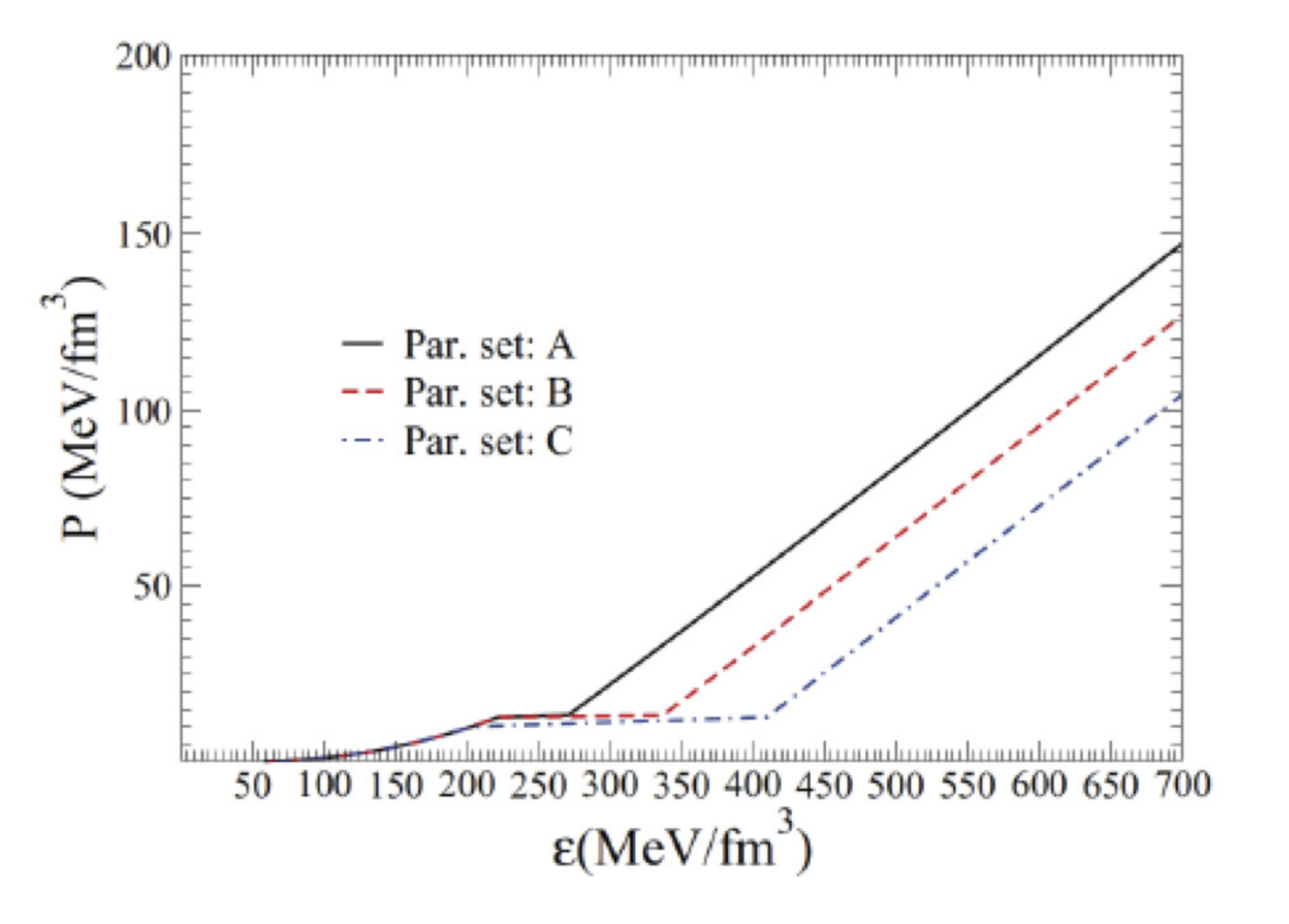}}
\vspace*{5pt}
\caption{Equation of state for hybrid star matter. Results for different bag pressures and interaction strengths are shown. }
\label{MITeos}
\end{figure}

At some unknown high density the system does not consist anymore of hadrons but of quarks. The main, unsolved question is when the transition to quark matter takes place. It is in principle quite possible that quarks only occur at densities larger than 6 to 10 times nuclear matter saturation density, which would imply that no neutron stars with a quark core, i. e. hybrid stars, would exist.
In a simplified study we determine the general impact of a quark core. For that purpose we adopt a standard equation of state for the hadrons (G300 \cite{G300}) and a MIT bag model equation of state for the quarks. 
For the latter EoS we vary the bag constant \cite{arXiv:1011.2233,Negreiros:2010tf}. The results are shown in the left panel of Fig. \ref{MIT}. One can clearly see that the onset of the population of quarks, signaled by the sharp bend in the curves, leads either to reduced
maximum masses of the stars or even to unstable solutions for the whole hybrid star branch. These calculations were done assuming non-interacting quarks in the quark core. Taking into account perturbative corrections due to the quark-quark interaction, the EoS of the quarks stiffens, stabilizing the hybrid star solutions (see Fig. \ref{MIT}, right panel). Changing the coupling constant to higher values stiffens the equation of state and the maximum star mass increases.
The equations of state corresponding to these results can be seen in the right panel of Fig. \ref{MITeos}. In case that the first-order phase transition to quark matter generates a large jump in energy density, the masses of the compact stars are reduced.
For a recent, even more schematic discussion of hybrid star properties assuming a simplified structure of the quark EoS, see \cite{Alford:2013aca}.

\begin{figure}[th]
\centerline{\includegraphics[width=9cm]{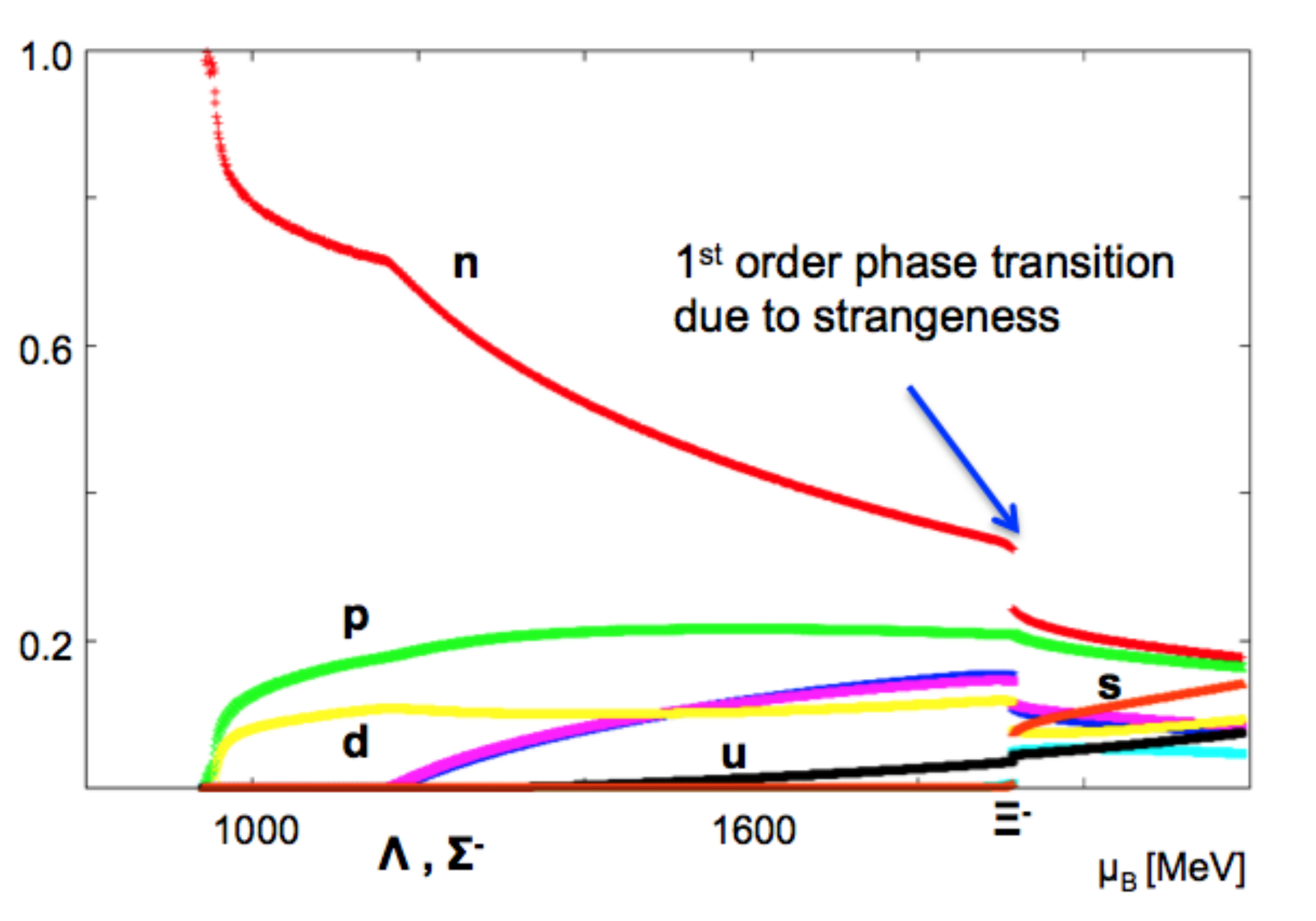}}
\vspace*{5pt}
\caption{Baryon number densities of baryons and quarks inside of the hybrid star. Quarks already enter the system a long time before the first-order phase transition that leads to a jump in the s-quark densities.}
\label{cocktail}
\end{figure}

\begin{figure}[th]
\centerline{\includegraphics[width=9cm]{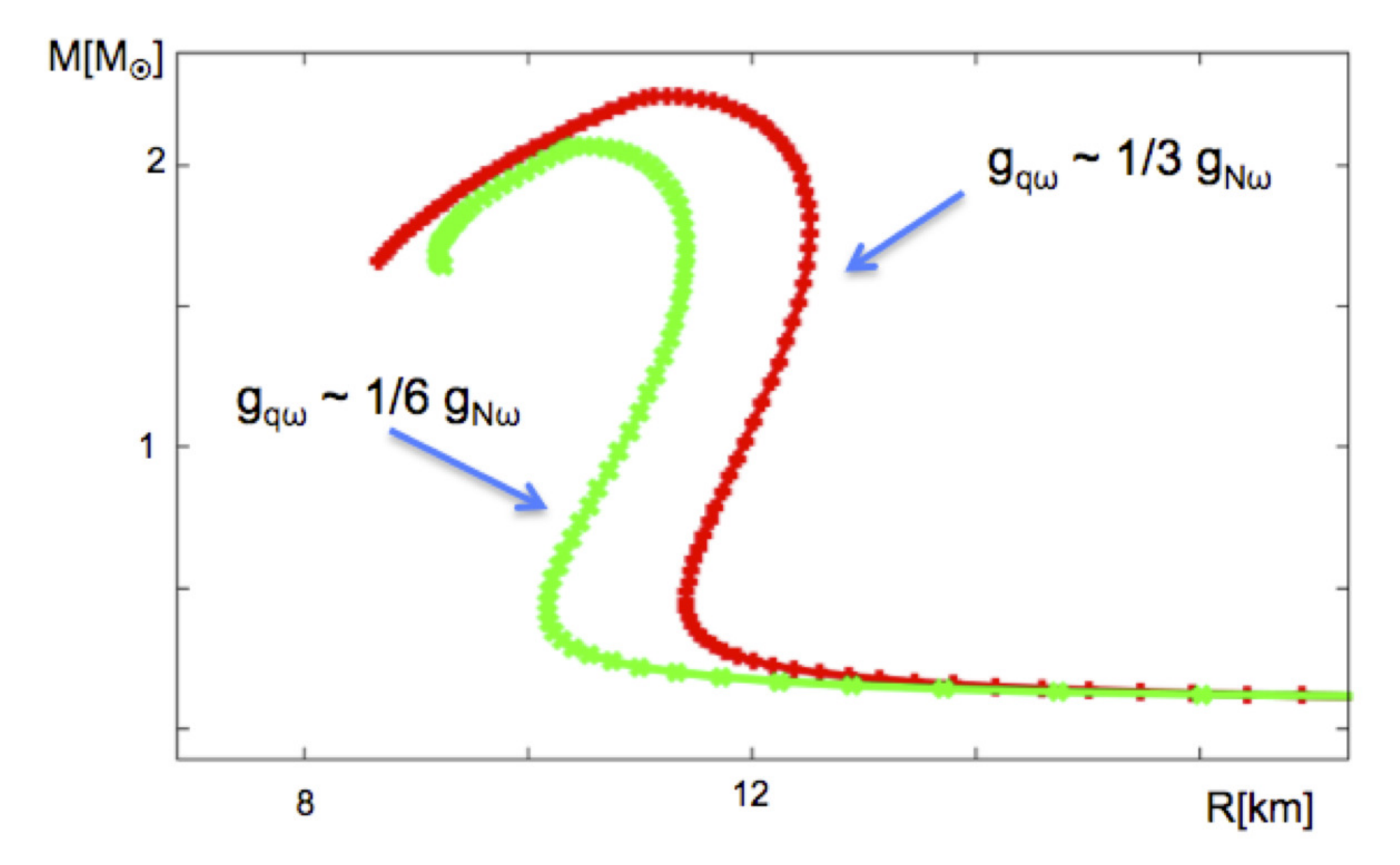}}
\vspace*{5pt}
\caption{Mass-radius diagram for compact stars within the quark-hadron (QH) model. The first-order transition causes a break for stable star solutions. Two curves for different choices of the quark vector-coupling are shown.}
\label{hybridstar}
\end{figure}

A more realistic description of a quark phase can be obtained by coupling quarks to the system in a way very analogous to the PNJL model by introducing a Polyakov loop field
$\Phi$ that describes the deconfinement phase transition \cite{Fukushima:2003fw,Ratti:2005jh}. In this case, the quarks couple to the mean fields in the same form as the hadrons in Eqs. (\ref{mass},\ref{cp}) with their own SU(3)-invariant
coupling strengths, thus coupling hadrons and quarks in a unified description \cite{Steinheimer:2010ib}.
While the isospin-symmetric matter does not show any first-order phase transition at high densities (see \cite{Steinheimer:2010ib}), star matter, i.e. charge neutral matter in $\beta$ equilibrium, does. The latter transition is generated by the strange quark sector and is therefore quite parameter-dependent and particularly sensitive to the coupling of the strange quark to the $\phi$ vector field. If one follows the baryon number densities contained in quark and hadron degrees of freedom, one can observe a very early onset of a mixed quark-hadron phase long before the first-order phase transition.
The particle densities of star matter can be seen in Fig. \ref{cocktail}. As the plot shows the baryon number densities, the corresponding quark densities have been multiplied by their baryon number $1/3$. Here one can observe the jump of the strange quark density at the critical chemical potential.
In Fig. \ref{hybridstar} the mass-radius diagram for this approach can be seen with star masses of up to $\sim 2.2 M_\odot$. 
As in the previous schematic studies with the MIT bag model repulsive vector interactions in the quark sector have been used \cite{Schramm:2013rya}. The results for different choices of the value are shown in the figure.
 
\begin{figure}[th]
\centerline{\includegraphics[width=9cm]{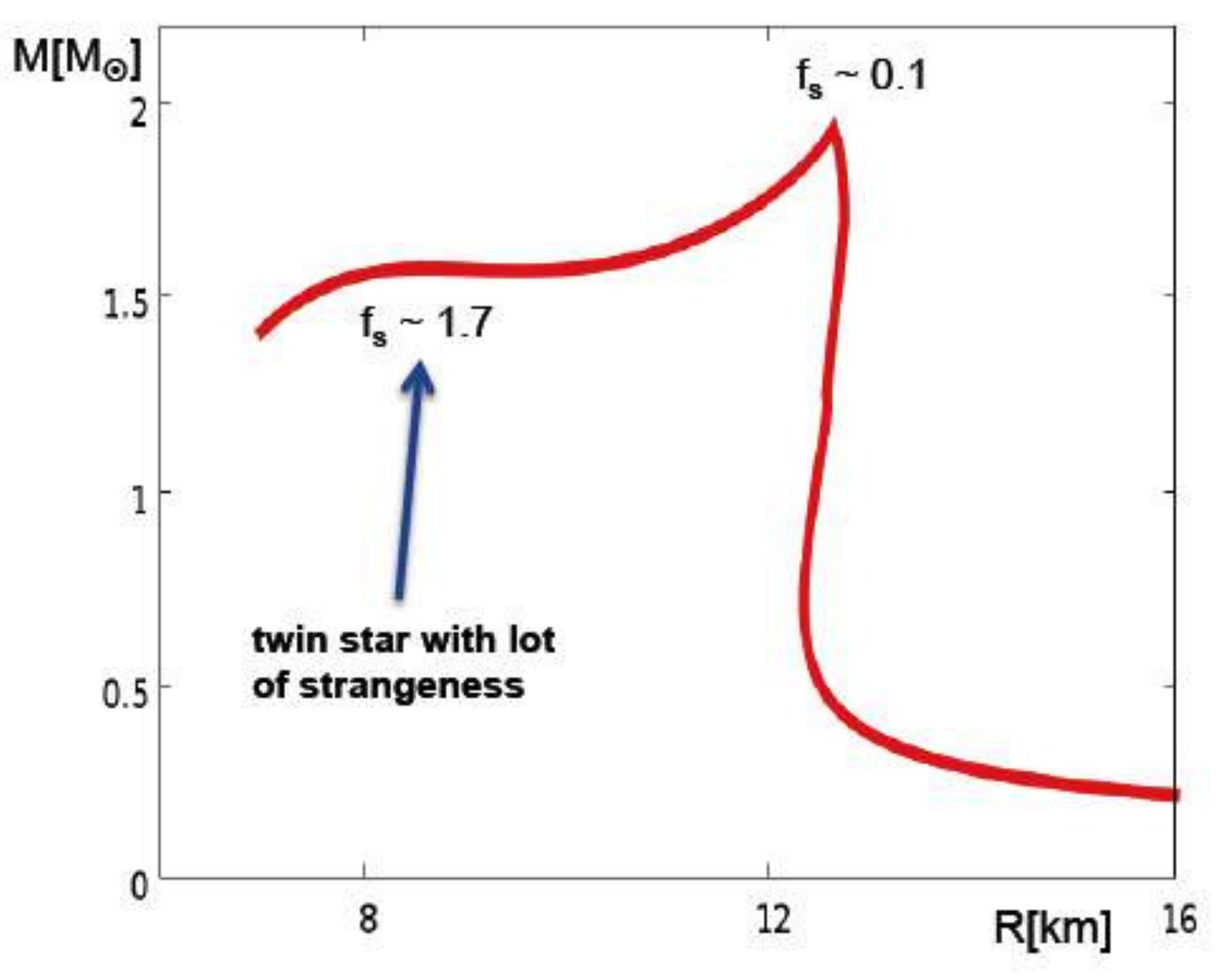}}
\vspace*{5pt}
\caption{Mass-radius diagram for compact stars for specific fine-tuned parameters. A strangeness-rich stable twin-star solution appears.}
\label{twin}
\end{figure}

For very specific parameter choices of a strong vector coupling of the strange quark the mass-radius diagram shows a second family of stable compact stars with a large strangeness content of values of strange quarks per baryon number $f_s$ larger than 1, and a distinctly smaller radius than the other stable branch, which also consists of hybrid stars albeit with very small strangeness content (Fig. \ref{twin}).

Thus, the general solution for obtaining a sizable quark core in a hybrid stars seems to be to introduce a repulsive quark vector interaction that stiffens the quark equation of state.
Note, however, that such an addition to models that contain quarks generates problems at the other end of the parameter space, i. e., at small chemical potential $\mu_B$ and high temperatures. In this case, lattice QCD simulations 
have determined the general behavior of the pressure of matter for small $\mu_B$ by calculating the Taylor coefficients for an expansion of the pressure in the chemical potential
\begin{equation}
 P = P(\mu_q=0,T) + c_2(T) \mu_q^2 T^2 + \ldots ~~ ,
 \end{equation}
 with the quark chemical potential $\mu_q = \mu_B / 3$. 
\begin{figure}[th]
\centerline{\includegraphics[width=6cm]{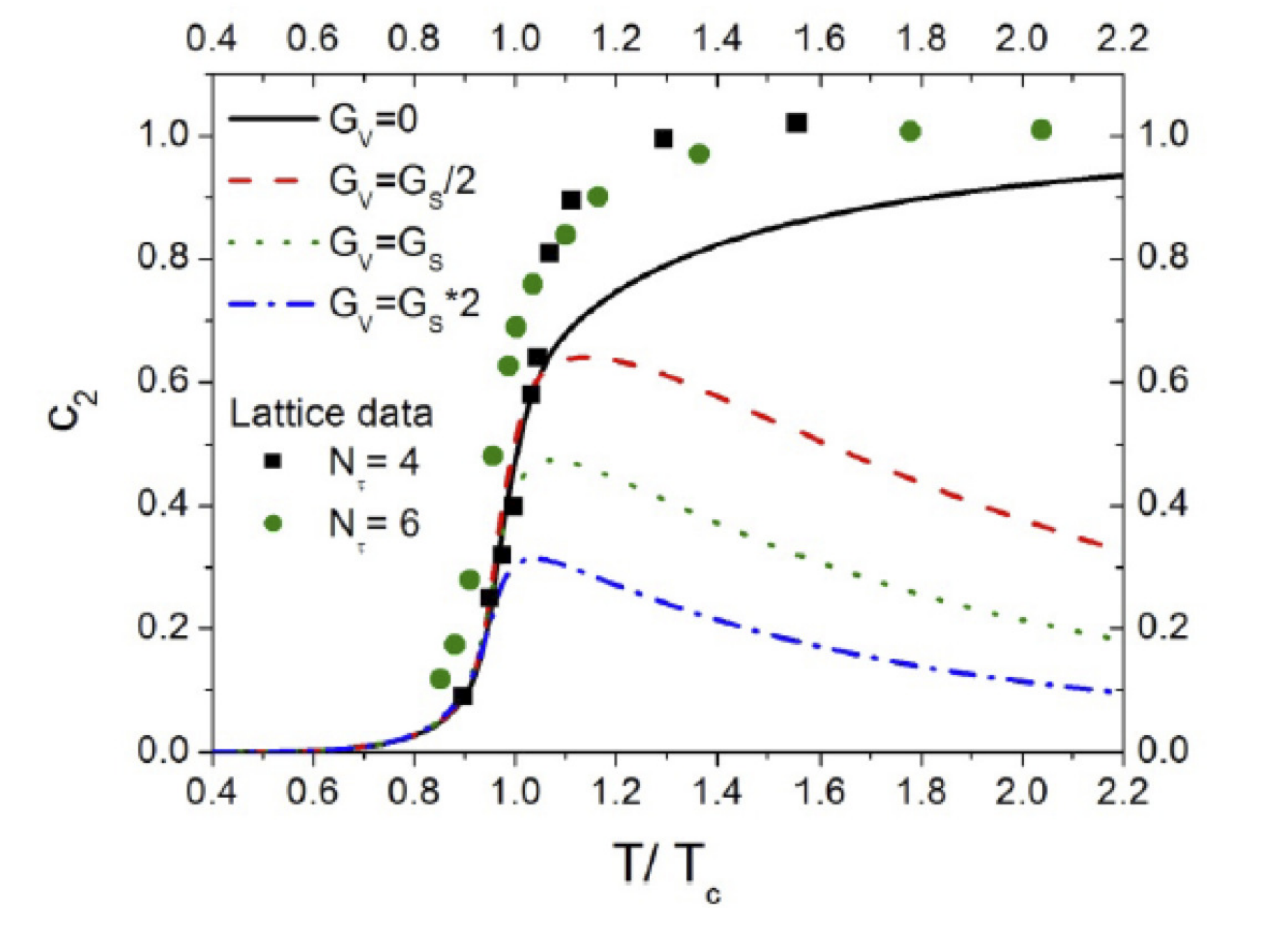}\includegraphics[width=6cm]{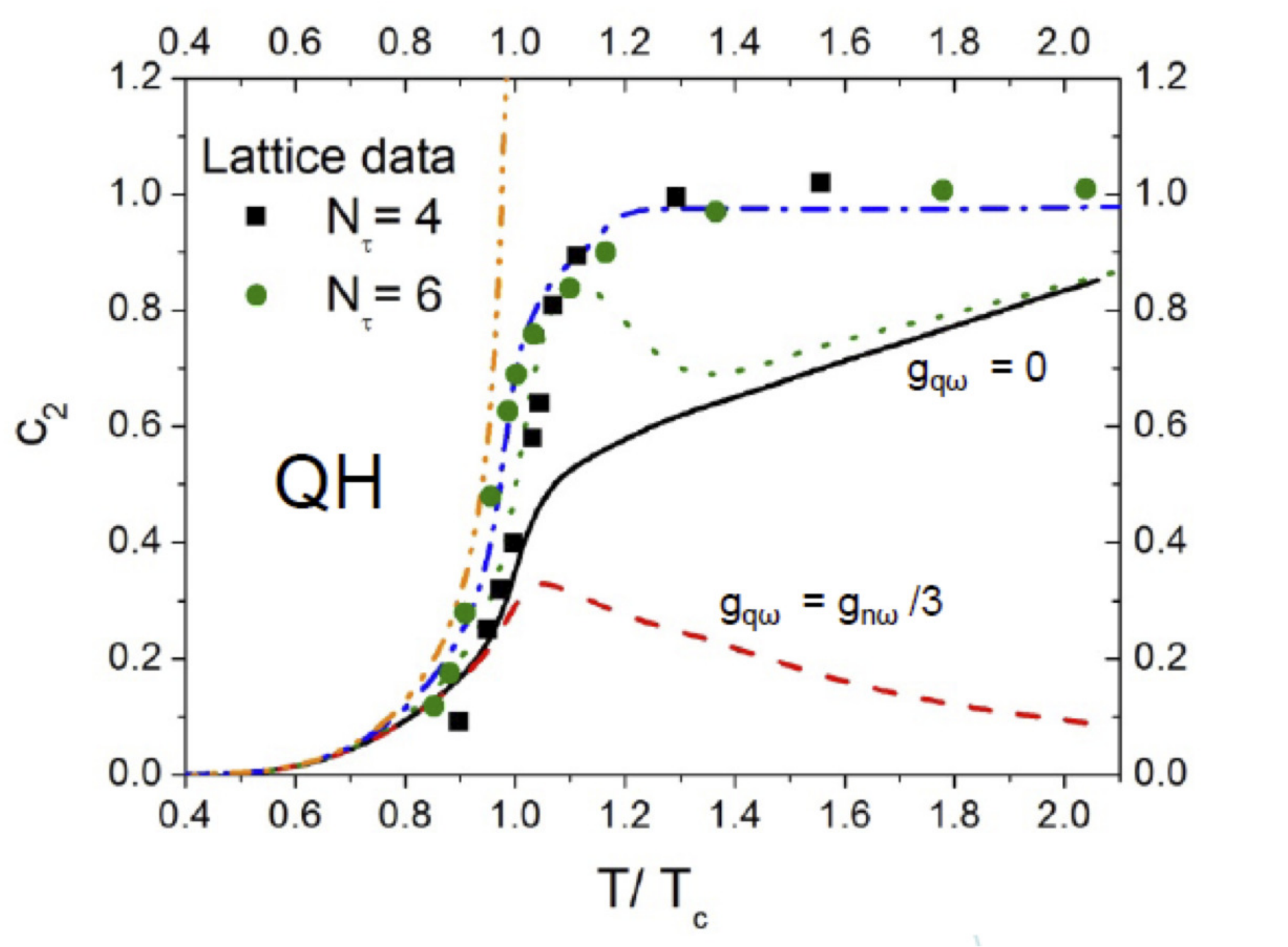}}
\vspace*{5pt}
\caption{Taylor coefficient $c_2$ of the expansion of the pressure in terms of chemical potential over temperature. Lattice QCD results are indicated by circles and squares. The lines show the results for different quark vector coupling.
The left panel displays the results for the PNJL model, whereas the right panel contains the computed values for the quark-hadron model \protect\cite{arXiv:1005.1176}.}
\label{c2}
\end{figure}
The panels in Fig. \ref{c2} show the results for zero and finite quark-vector coupling for different model approaches. The results indicate that, independent of the specific model applied, the curves with a non-zero coupling strength deviate significantly from lattice data \cite{Bazavov:2012jq,Cheng:2008zh,Borsanyi:2011sw}, far beyond any statistical or systematic error bars \cite{arXiv:1005.1176,Rau:2013xya}. Therefore, there is a currently unresolved problem with describing neutron stars that contain a  sizable quark core and simultaneously achieving agreement with lattice data.  

\section{Conclusions}
\label{sec:5}

We have discussed compact star properties and the impact of exotic particles inside stars on the attainable maximum star masses.
We have shown that in a reasonable flavor-SU(3) approach a 2-solar mass star, a so-called hyper star, can be obtained, even when one includes effects of the $\Delta$ resonances.
However, the amount of strangeness contained in the star through the population of hyperons is rather small.
Considering a kaon condensate we have shown that the critical density for the onset of condensation in our model is too high to have any impact on star properties.
If one extends the approach to consider quarks in the inner regions of the star (a hybrid star), we have demonstrated that a simple addition of a quark phase via a non-interacting MIT model leads
to either strongly reduced masses or no stable hybrid star solutions at all. Interactions between the quarks that produce a stiffening of the equation of state can remedy this effect.
In the generalized quark-hadron model, a heavy hybrid star of 2.2 solar masses is obtained without any phase mixing (as obtained in the usual Gibbs construction), as the quarks co-exist with hadrons without any first-order transition.
A later phase transition is triggered by a large increase of the strange quark density. This value, however, is strongly parameter-dependent.

Finally we discussed the general problem with adopting a large quark-vector coupling in order to stiffen the quark equation of state and, in turn, obtain high-mass hybrid stars.
Comparing results for small chemical potential with QCD lattice simulations shows a clear discrepancy generated by these interactions for very different model approaches.
Thus, a consistent picture of a heavy hybrid star and agreement with lattice date is still elusive.

\begin{acknowledgement}
We acknowledge the use of the CSC computer facilities at Frankfurt university for our work. T. S. is supported by the Nuclear Astrophysics Virtual Institute (NAVI).
\end{acknowledgement}

\end{document}